\documentclass[10pt,letterpaper,compsoc,conference]{iiswc26}

\usepackage{cite}
\usepackage{amsmath,amssymb,amsfonts}
\usepackage{graphicx}
\usepackage[dvipsnames]{xcolor}
\usepackage[final]{microtype}
\usepackage[italic]{mathastext}
\usepackage[T1]{fontenc}
\usepackage{textcomp}
\usepackage[all]{nowidow}
\usepackage{flushend}
\usepackage{enumitem}

\usepackage{comment}

\usepackage{tikz}
\newcommand{\circnum}[1]{%
\tikz[baseline=(char.base)]{
\node[
    shape=circle,
    fill=black,
    text=white,
    inner sep=1pt,
    font=\scriptsize
] (char) {#1};}}

\usepackage{booktabs}
\usepackage{multirow}
\usepackage{array}

\graphicspath{{figs/}}
\newcommand{\figwidth}{\columnwidth}

\IEEEoverridecommandlockouts

\begin{document}

\title{Dead on Arrival: Characterizing and Protecting Against Dead-Entry TLB Misses in GPU Microarchitectures}

\author{
\IEEEauthorblockN{Shafayat Mowla Anik\IEEEauthorrefmark{1},
Yongchan Jung\IEEEauthorrefmark{2},
Jeeho Ryoo\IEEEauthorrefmark{2},
Byeong Kil Lee\IEEEauthorrefmark{1}}
\IEEEauthorblockA{\IEEEauthorrefmark{1}University of Colorado at Colorado Springs, Colorado Springs, CO\\
Email: \{sanik, blee\}@uccs.edu}
\IEEEauthorblockA{\IEEEauthorrefmark{2}Fairleigh Dickinson University, Vancouver, BC, Canada\\
Email: y.jung@student.fdu.edu, j.ryoo@fdu.edu}
\thanks{This work has been submitted to the IEEE for possible publication. Copyright may be transferred without notice, after which this version may no longer be accessible.}
}

\maketitle
\pagestyle{plain}

\begin{abstract}
GPU workloads with large memory footprints frequently suffer from redundant L2 TLB misses
in which a recently evicted translation is immediately re-walked at full page-walk cost.
We characterize these \emph{dead-entry misses} across 24 GPU workloads, finding they
account for up to 99\% of L2 TLB misses in the most TLB-sensitive applications, yet their
performance impact varies widely depending on memory access structure.
Workloads where warps share the same virtual page suffer from burst amplification, where
a single eviction stalls many warps simultaneously waiting for one translation to return.
In contrast, workloads where each warp accesses a distinct set of pages face a
capacity-overflow problem that no replacement policy can resolve, a distinction validated
by huge page experiments.
Building on this two-class taxonomy, we design \emph{DEPOT} (Dead-Entry PrOTection),
a 1\,KB Bloom filter mechanism that prevents recently evicted translations from being
displaced immediately upon reinstallation, delivering up to 72\% IPC improvement on
interference-driven workloads with zero overhead on others, and composing with the state-of-the-art TLB prefetching and compaction mechanism,
for 2 to 7\% additional gain.
\end{abstract}

\section{Introduction}
\label{sec:introduction}

Address translation is a first-order performance concern in modern GPU workloads~\cite{power_x86}.
As applications move toward larger, irregular memory footprints such as sparse graph
traversal, scientific stencils, and machine learning inference over large embedding tables,
the GPU L2 TLB becomes a sustained bottleneck \cite{ganguly_interplay, shin_irregular}.
Prior work has approached this problem almost exclusively from an access-pattern perspective,
predicting which virtual pages will be needed next and prefetching their translations
before the demand miss occurs~\cite{latpc,valkyrie, mosaic, snake_prefetch}.
This framing implicitly treats each TLB miss as a cold event requiring a fresh page walk,
but many TLB misses are not cold at all. A \emph{dead-entry TLB miss} occurs when a virtual page whose translation is still valid
is evicted from the TLB and then re-accessed before the working set rotates again.
The page walk is not cold; it is redundant.
The entry was alive, was discarded by LRU replacement, and is now being reconstructed
at full page-walk cost.
Measuring this phenomenon across 24 GPU workloads reveals that dead-entry misses are
pervasive: in the most TLB-sensitive applications, over 99\% of all L2 TLB misses
carry an eviction-history signature.

\begin{figure}[t]
  \centering
  \includegraphics[width=\figwidth]{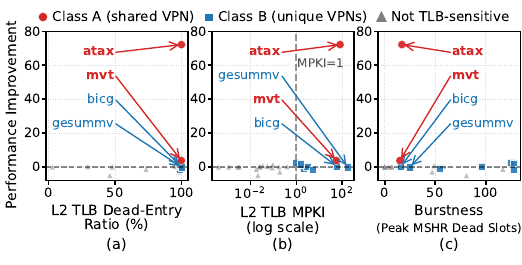}
  \caption{Performance improvement (IPC) from eliminating dead-entry re-walks against three candidate predictors: (a)~DE ratio, (b)~MPKI, and (c)~MSHR burstiness, for all 24 workloads.}
  \label{fig:teaser}
\end{figure}

Yet the performance consequences of dead-entry misses are not uniform, and
Figure~\ref{fig:teaser} shows why simple miss statistics fail to explain the divergence.
Each plot shows the performance improvement from eliminating dead-entry re-walks against a different
candidate predictor for all 24 workloads.
Figure~\ref{fig:teaser}(a) shows that dead-entry ratio has no predictive power: all TLB-sensitive workloads
cluster near 99\% dead-entry ratio, yet their performance improvement ranges from near zero to $72\%$.
Figure~\ref{fig:teaser}(b) shows that MPKI is necessary but not sufficient: workloads with large IPC improvements
and workloads with near-zero improvements remain indistinguishable at similar MPKI levels.
Figure~\ref{fig:teaser}(c) reveals the discriminator: burstiness, measuring how many pending translation requests
are collectively served by a single page walk; workloads with high burstiness achieve large performance
improvements, while those with near-zero burstiness achieve near-zero improvements.

In some workloads, all threads in a warp reference the same memory page, so a single TLB
eviction stalls many warps simultaneously waiting for one translation to return.
Eliminating that redundant re-walk unblocks all of them at once, producing large performance
improvements and high observed burstiness.
In other workloads, each thread accesses a distinct page from a working set that exceeds
TLB capacity.
Evictions are driven by the sheer size of the working set, and no replacement-level
mechanism can eliminate them because the TLB simply cannot hold all required translations
at once.
These two root causes lead to opposite outcomes under any dead-entry protection mechanism,
and we characterize and validate both across all 24 workloads in Section~\ref{sec:characterization}.

Building on this characterization, we design \emph{DEPOT} (Dead-Entry PrOTection),
a lightweight hardware mechanism that targets the first type of workload.
DEPOT uses a small Bloom filter to track recently evicted TLB entries and temporarily
protects re-installed entries from being displaced again, preventing the same redundant
re-walks from compounding.
The mechanism requires approximately 3.5\,KB of storage, introduces no prediction logic,
and falls back gracefully to standard LRU replacement for workloads where it provides
no benefit.
DEPOT also composes well with LatPC~\cite{latpc}, the state-of-the-art TLB prefetching
and compaction mechanism, because the two techniques address fundamentally different
sources of misses.
LatPC reduces misses through stride prediction and entry compaction; DEPOT prevents
recently used pages from being discarded prematurely.
Together they outperform either mechanism alone by 2 to 7\% on TLB-sensitive workloads
at negligible additional hardware cost.

In this paper, we make the following contributions.
\begin{itemize}
  \item \textbf{Eviction-history characterization.}
    We measure GPU L2 TLB misses by eviction history across 24 workloads,
    establishing dead-entry re-walks as a prevalent, structurally distinct miss class.
  \item \textbf{Two-class taxonomy.}
    We identify two root causes among TLB-sensitive workloads,
    interference-driven misses where multiple warps share the same virtual page,
    and capacity-driven misses where the working set exceeds TLB reach,
    and validate the taxonomy using 2\,MB huge page experiments.
  \item \textbf{DEPOT mechanism and composition.}
    We design DEPOT (Dead-Entry PrOTection), a Bloom-filter mechanism
    (1\,KB filter, 3.5\,KB total)
    achieving up to 72\% performance improvement on interference-driven workloads and
    2 to 7\% additive improvement on top of LatPC~\cite{latpc},
    the state-of-the-art TLB prefetching and compaction mechanism.
\end{itemize}

The remainder of this paper is organized as follows.
Section~\ref{sec:background} reviews GPU address translation and dead-entry TLB misses.
Section~\ref{sec:related} summarizes related work.
Section~\ref{sec:methodology} describes the methodology and experimental setup.
Section~\ref{sec:characterization} presents the dead-entry miss characterization and workload taxonomy.
Section~\ref{sec:depot} introduces DEPOT, and Section~\ref{sec:evaluation} evaluates its performance and composition with the state-of-the-art scheme.
Section~\ref{sec:discussion} discusses limitations, and Section~\ref{sec:conclusion} concludes.
\section{Background}
\label{sec:background}

\subsection{GPU Address Translation}

Modern GPUs execute code in units called \emph{warps}, groups of 32 threads that advance
in lock-step~\cite{aamodt_book,che_gpgpu_study,garland_cuda}.
Every memory instruction in a warp generates up to 32 virtual addresses simultaneously,
and before any of those accesses can reach the cache hierarchy, each address must be
translated by mapping its virtual page number (VPN) to a physical frame number (PFN)
through the TLB.
Unlike CPUs where a single thread issues one translation at a time, a GPU warp can
present up to 32 distinct VPNs per cycle, creating significant pressure on the
translation hardware.
In practice, coalescing~\cite{nyland_coalescing} reduces this number when multiple threads address the same page,
but memory-intensive workloads with large or strided access patterns routinely generate
one or more unique VPNs per warp instruction.

\begin{figure}[t]
  \centering
  \includegraphics[width=\figwidth]{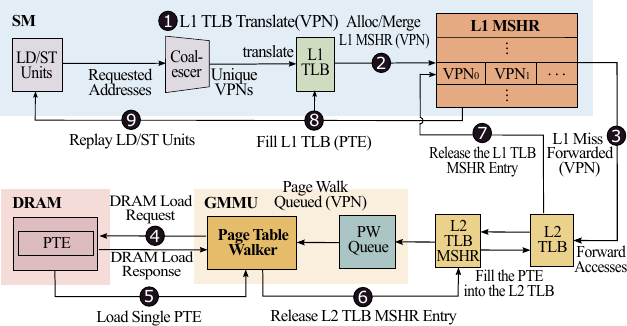}
  \caption{GPU address translation hierarchy (SM86): L1 TLB per SM, shared L2 TLB,
  16 PTWs, and MSHR merging for same-VPN requests.}
  \label{fig:arch}
\end{figure}

Figure~\ref{fig:arch} traces the full translation pipeline for a warp-level memory
instruction.
The LD/ST units generate virtual addresses that pass through the coalescer, which
extracts unique VPNs and presents them to the per-SM L1 TLB for lookup~\circnum{1}.
Each Streaming Multiprocessor (SM) hosts a private, fully-associative L1 TLB holding
32 entries~\cite{pichai_gpu_translation,translation_cpu}; on the Ampere SM86 microarchitecture~\cite{nvidia_ampere} studied here, 46 SMs operate in
parallel, each with its own independent L1 TLB, flushed at every kernel boundary.
An L1 hit resolves the translation in a single cycle.
On an L1 miss, the VPN is allocated into the L1 MSHR \circnum{2}: if a second request arrives
for the same VPN while the first is still in flight, it is merged into the existing
MSHR entry rather than generating a duplicate request, and both requestors are
unblocked together when the translation returns.
A VPN that finds no L1 MSHR entry already in flight is forwarded to the shared L2
TLB \circnum{3}. The L2 TLB is chip-wide, holding 1024 entries on SM86, and services all 46 SMs
concurrently. With 4\,KB pages and 1024 entries, the total TLB reach spans
$1024 \times 4\,\text{KB} = 4\,\text{MB}$ of virtual address space; workloads whose
active footprints exceed this threshold will encounter L2 misses regardless of
replacement policy.
On an L2 miss, the L2 TLB performs the same MSHR merge as the L1: same-VPN requests
are coalesced so that one page-table walk serves all waiting warps.
The VPN is then dispatched through the page walk queue to one of the 16 dedicated
hardware page-table walkers (PTWs)~\cite{shin_irregular}, each of which traverses the GPU's hierarchical
page table by issuing sequential reads to DRAM~\circnum{4}.
The GPU memory management unit (GMMU) orchestrates these page walks following an L2 TLB miss.
A single walk incurs multiple DRAM round trips, with total latency on the order of
hundreds to over a thousand cycles depending on DRAM occupancy and row-buffer state.
Because the GPU has only 16 PTWs, it can service at most 16 walks concurrently;
newly arriving L2 misses queue behind occupied PTWs, adding queuing delay on top
of the walk itself \circnum{5}.
When the PTW finishes, the PTE arrives back from DRAM \circnum{5} and the L2 TLB MSHR entry
is released, unblocking all warps that were waiting on that VPN at the L2 level \circnum{6}.
The translation then propagates back up: the L1 TLB MSHR entry is released \circnum{7},
the PTE is installed into the L1 TLB \circnum{8}, and the stalled warp re-executes its
memory instruction from the LD/ST units \circnum{9}.

The performance significance of this pipeline lies in the compounding of stall latencies.
A warp stalls at its L1 TLB and cannot issue further instructions until the translation
returns.
An L2 miss extends that stall by the full page-walk duration, typically hundreds of
cycles.
More importantly, when multiple warps reference the same VPN simultaneously, a single
eviction can stall many warps at once through the L2 MSHR merge mechanism, making
a single dead-entry event far more costly than the raw walk latency suggests.
This PTW bottleneck is the primary reason TLB miss behavior dominates performance in
translation-sensitive GPU workloads~\cite{pichai_gpu_translation,avoiding_tlb,latpc}.

\subsection{Dead-Entry TLB Misses}

We define a \emph{dead-entry TLB miss} formally as follows.

\begin{center}
\fbox{\begin{minipage}{0.92\columnwidth}
\smallskip
\emph{A TLB miss at virtual page $p$ at time $t$ is a \textbf{dead-entry miss} if $p$
was previously installed in the TLB, subsequently evicted before time $t$, and the
eviction preceded the miss without any intervening access to $p$ that would have
re-established a valid entry.}
\smallskip
\end{minipage}}
\end{center}

\noindent Equivalently, a dead-entry miss is one where the page-table walker reconstructs a
translation that was valid and present in the TLB within the current execution window,
then discarded by the LRU replacement policy~\cite{lru,belady}.
The walk is logically redundant: the hardware performs it because the eviction destroyed
the cached result, not because the translation itself is new or stale.

It is important to understand how this definition relates to the classical miss taxonomy.
The traditional three-C classification (compulsory, capacity, and conflict)~\cite{hill_smith_3c} describes
why an eviction occurred.
Dead-entry is an orthogonal characterization that describes what the evicted entry was,
regardless of what caused the eviction.
A dead-entry miss can arise from a capacity eviction, where the working set exceeds TLB
reach, or from an interference-driven eviction, where a burst of other pages displaces
a still-needed entry.
These two root causes correspond to the two workload classes identified in
Section~\ref{sec:characterization}.
Crucially, not every capacity or conflict eviction produces a dead-entry miss: if the
evicted page is never accessed again within the current execution window, there is no
re-walk and no dead-entry event.

The orthogonality matters because it changes what is actionable.
A cold miss cannot be avoided without prediction, as the hardware has never seen the page.
A capacity miss cannot be avoided without more TLB entries or larger pages.
A dead-entry miss, by contrast, is structurally redundant: the translation is unchanged,
so no new information is gained by re-walking the page table.
It is avoidable in principle simply by remembering that the entry was recently evicted
and protecting it from immediate re-eviction.
This is the observation that motivates DEPOT.

\subsection{MSHR Merging}

The L2 TLB MSHR tracks in-flight page walks and coalesces requests for the same VPN.
When a second warp issues a miss for a VPN already under walk, it is merged into the
existing MSHR entry rather than issuing a new walk, so one walk satisfies all merged
requests simultaneously.
The consequence for dead-entry misses is amplification.
In workloads where many warps simultaneously reference the same VPN, when that VPN
is evicted and re-walked, the walk serves a burst of coalesced requests.
The MSHR dead-slot occupancy spikes sharply during the walk and drains when the fill
completes.
This burst structure is the mechanism by which a single dead-entry event produces
disproportionately large IPC degradation, and by which protecting a single re-installed
entry produces disproportionately large IPC recovery.

\section{Related Work}
\label{sec:related}

\noindent\textbf{TLB prefetching and compaction.}
LatPC~\cite{latpc} combines a per-entry stride predictor-building on classic
distance-based TLB prefetching~\cite{kandiraju_distance}-with a compaction mechanism
that consolidates multiple TLB entries covering a contiguous address range, issuing
prefetch requests before demand misses occur by learning access patterns from prior hits.
Valkyrie~\cite{valkyrie} extends stride-based prefetching with multi-level stride
detection and cooperative prefetching across SMs.
Avatar~\cite{avatar} prefetches translations across address-space boundaries to support
multi-tenant workloads.
More recently, SnakeByte~\cite{snakebyte} introduces an adaptive and recursive
page-merging design that grows TLB entries dynamically based on access locality,
and STAR~\cite{star,li_sharing_spilling} exploits sub-entry sharing to improve TLB efficiency on
multi-instance GPUs.
On the CPU side, Bhattacharjee~\cite{bhattacharjee_ttp,bhattacharjee_large} proposes
translation-triggered prefetching that initiates speculative TLB fills upon
preceding translations, and Margaritov et al.~\cite{margaritov_pat,margaritov_learned} reduce
TLB-miss latency via prefetched address translation that pipelines walks ahead
of demand.
All of these mechanisms operate on the access sequence, predicting future translations
from observed access history.
Our work instead characterizes misses by eviction history, a perspective orthogonal
to access-pattern prediction.

\noindent\textbf{Page-table walk optimization.}
A second line of recent work accelerates the cost of translations that do miss the
TLB hierarchy.
R2D2~\cite{r2d2} eliminates redundant page walks by exploiting linearity in the
addresses generated across the threads of a warp.
Heliostat~\cite{heliostat} repurposes ray-tracing accelerators to perform multiple
page-table walks in parallel.
Trans-FW~\cite{trans_fw} short-circuits walks in multi-GPU systems via remote
forwarding, and Barre Chord~\cite{barre_chord} addresses translation efficiency for
multi-chip-module GPUs.
Earlier work in the speculative-walk space includes SpecTLB~\cite{pham_spectlb,barr_translation_caching},
which executes ahead of confirmed translations to overlap walk latency with useful
work, and Griffin~\cite{baruah_griffin}, which provides hardware-software support for
efficient page migration in multi-GPU systems.
These mechanisms optimize the cost of walks that occur; our characterization
identifies a complementary opportunity: the dead-entry re-walk class, where the
walk itself is redundant because the entry was recently in the TLB.

\noindent\textbf{Page-table layouts and huge pages.}
Increasing the page size from 4\,KB to 2\,MB expands TLB reach by a factor of 512
without adding entries.
Mosaic~\cite{mosaic} proposes a hardware range TLB that achieves similar reach without
requiring OS support or physically contiguous memory.
CoLT~\cite{colt} uses contiguous TLB entries that cover small address ranges at 4\,KB
granularity.
Elastic Cuckoo Page Tables~\cite{elastic_cuckoo} restructure the page-table layout to
expose parallelism during translation, while Translation Ranger~\cite{translation_ranger,karakostas_rmm}
provides OS support for contiguity-aware TLBs.
Pham et al.~\cite{pham_clustering,park_hybrid} extend TLB reach by exploiting natural clustering in
page translations, capturing partial contiguity without the full alignment requirements
of huge pages.
These mechanisms address the fundamental capacity limitation of small-page TLBs.
Our work uses 2\,MB huge pages as a diagnostic tool to distinguish capacity-driven miss
patterns from interference-driven ones, not as a deployment target.

\noindent\textbf{TLB-aware scheduling and concurrency.}
Pichai et al.~\cite{cta_schedule} reduce TLB pressure by grouping cooperative thread
arrays with overlapping virtual address footprints onto the same SM, improving L1 TLB
locality and lowering the miss rate.
MASK~\cite{avoiding_tlb,pichai_gpu_translation} redesigns the GPU memory hierarchy to support multi-application
concurrency, addressing TLB shootdown and coherence overheads.
OASIS~\cite{oasis} introduces object-aware page management for multi-GPU systems,
placing memory closer to the GPU that touches it most frequently.
Kim et al.~\cite{kim_batch,shin_irregular} propose batch-aware unified memory management for
irregular GPU workloads, coalescing page-management requests to amortize translation
overhead.
These techniques reduce the frequency of TLB misses through workload-aware scheduling
and placement.
Our characterization focuses on the structural cause of misses that persist regardless
of scheduling, specifically the re-walk of recently evicted entries.

\noindent\textbf{Dead-block and dead-page prediction.}
The dead-entry concept has a long lineage in cache replacement~\cite{rrip,hawkeye,victim_cache}.
Khan et al.~\cite{sdbp_micro10} introduced sampling-based dead-block prediction for
last-level caches, identifying blocks unlikely to be reused before eviction and
prioritizing them for replacement.
Most closely related to our work, Mazumdar et al.~\cite{dop_hpca21} propose joint
dead-page and dead-block predictors that clean TLBs and caches together using
PC-based reuse prediction.
Our characterization identifies a structurally distinct dead-entry re-walk
miss class in GPU L2 TLBs, where eviction history rather than access-PC reuse is
the discriminating signal, and quantifies its prevalence across 24 workloads.

\noindent\textbf{MSHR management and memory-level parallelism.}
Prior work on adaptive MSHR sizing~\cite{mshr_mgmt} and memory-level parallelism
control~\cite{mlp} has shown that MSHR occupancy patterns carry workload-classification
signal useful for tuning cache and memory behavior.
We apply an analogous observation to the TLB MSHR, showing that the temporal shape
of MSHR dead-slot occupancy distinguishes structurally different classes of
TLB-sensitive workloads without offline profiling.
CPU TLB characterization~\cite{translation_cpu,shared_tlb,bhattacharjee_inter} identifies
working-set overflow as the dominant miss cause in server workloads; GPU address
translation differs in that thousands of concurrent warps interact through a shared L2
TLB and MSHR, creating amplification effects with no direct CPU analog.

\section{Methodology}
\label{sec:methodology}

\subsection{Simulation Platform}

All experiments run on Accel-Sim~\cite{accelsim}, a cycle-accurate GPU simulator
built on GPGPU-Sim~\cite{gpgpusim}.
We adopt Accel-Sim's default \texttt{SM86\_RTX3070} configuration~\cite{accelsim},
which models an NVIDIA GeForce RTX~3070-class GPU~\cite{rtx3070} based on the Ampere
microarchitecture~\cite{nvidia_ampere,ampere_microbench}; the baseline GPU, cache, and
memory parameters are used unmodified.
The TLB hierarchy parameters---L1 and L2 TLB capacities, MSHR depths, lookup latencies,
and the 16 page-table walkers---follow the configuration used in the state-of-the-art work~\cite{latpc},
ensuring our results are directly comparable to that prior work.
Table~\ref{tab:tlb_config} summarizes the configuration.
The L2 TLB has 1024 entries with LRU replacement, providing 4\,MB of reach with 4\,KB pages.
Page-table walks are modeled with 16 dedicated PTWs, each capable of one concurrent walk.
The L2 MSHR supports 128 in-flight translation requests; same-VPN requests are merged
into a single MSHR entry.

\begin{table}[ht]
  \centering
  \caption{Simulated System Configuration.}
  \label{tab:tlb_config}
  \scriptsize
  \setlength{\tabcolsep}{4pt}
  \renewcommand{\arraystretch}{1.2}
  \begin{tabular}{|>{\centering\arraybackslash}m{0.14\columnwidth}|m{0.74\columnwidth}|}
    \hline
    \multicolumn{2}{|c|}{\textbf{GPU Configuration}} \\
    \hline
    SM & 46 SMs, 1536 threads / 32-thread warps, 32 CTAs per SM\newline
         65{,}536 registers per SM, up to 100\,KB shared memory\newline
         1132\,MHz core clock \\
    \hline
    Cache & 128\,B line; L1 D-cache 128\,KB, 32-way, 384 MSHRs\newline
            L2 cache (unified) 4\,MB, 16-way \\
    \hline
    DRAM & GDDR6, 16 channels, 14\,Gbps, 3500.5\,MHz\newline
           254-cycle access latency \\
    \hline
    \hline
    \multicolumn{2}{|c|}{\textbf{Virtual Memory Configuration}} \\
    \hline
    L1 TLB & 32 entries, fully associative, 4\,KB pages\newline
             16 MSHRs (4-way merge), 4 ports\newline
             20-cycle lookup latency, LRU replacement \\
    \hline
    L2 TLB & 1024 entries, 16-way, 4\,KB pages\newline
             128 MSHRs (8-way merge), 16 ports\newline
             80-cycle lookup latency, LRU replacement\newline
             Reach 4\,MB (4\,KB) / 256\,MB (2\,MB pages) \\
    \hline
    Page Walk & 16 page-table walkers\newline
                x86-64 4-level page table~\cite{amd64_manual,intel_sdm},
                254 cycles per level\newline
                32-entry page-walk cache, 20-cycle lookup \\
    \hline
  \end{tabular}
\end{table}

\subsection{Instrumentation}

We extend Accel-Sim with four measurement and mechanism modules:

\noindent\textbf{Miss arrival timeseries:} Records arrival timestamps of every
L2 TLB miss, enabling analysis of temporal miss structure.

\noindent\textbf{Dead-entry detection:} Tracks the last eviction cycle per VPN
in a hash table. A miss is classified as dead-entry if the VPN appears in this
table, indicating a prior eviction since last installation.

\noindent\textbf{MSHR dead-slot timeseries:} Samples the number of MSHR entries
currently waiting on dead-entry re-walks, at 100-cycle intervals.
Peak occupancy defines the \emph{burstiness} metric.

\noindent\textbf{DEPOT protection statistics:} Records Bloom filter insert and hit
counts, protected-entry counts, and LRU fallback events. More implementation details are in Section~\ref{sec:depot}.

\subsection{Workload Suite}
Table~\ref{tab:workloads} characterizes all 24 workloads
(Classes: \textbf{A}\,=\,interference-driven; \textbf{B}\,=\,capacity-driven; \textbf{---}\,=\,not TLB-sensitive). Here, MPKI is L2 TLB misses per kilo-instruction, while mem-MPKI is misses per
kilo memory-instruction, isolating translation pressure from non-memory compute.
We select applications with working sets exceeding 4\,MB or L2 TLB MPKI above 1.0,
matching the criteria used in LatPC~\cite{latpc}.
Workloads span Rodinia~\cite{rodinia}, PolyBench~\cite{polybench}, Lonestar~\cite{lonestar},
Parboil~\cite{parboil}, and Pannotia~\cite{pannotia} benchmark suites.
We apply a two-bin framing: \textbf{TLB-sensitive} workloads (L2 MPKI $\geq 1$, matching the
selection criterion of LatPC~\cite{latpc}) vs.\ \textbf{not TLB-sensitive} workloads
(L2 MPKI $< 1$).
Nine workloads meet the TLB-sensitive threshold (including \texttt{mst} at MPKI$\,{=}\,0.98$,
a near-threshold borderline case with confirmed capacity-driven root cause);
the remaining 15 serve as a graceful-degradation validation set.
Within the TLB-sensitive bin, we sub-classify by access structure:
\textbf{Class~A} (interference-driven) for workloads where all threads in a warp
reference the same virtual page (shared VPN)---producing MSHR-merge amplification---and
\textbf{Class~B} (capacity-driven) for workloads where each warp instruction generates
distinct VPNs per thread, driving working-set-overflow evictions that exceed TLB reach.
Class~B membership is confirmed by the 2\,MB IPC experiment: if switching to 2\,MB pages
(L2 reach $= 256$\,MB) yields a large IPC speedup, the workload was reach-limited.

\begin{table}[t]
  \centering
  \caption{List of Evaluated Workloads.}
  \label{tab:workloads}
  \scriptsize
  \setlength{\tabcolsep}{4pt}
  \renewcommand{\arraystretch}{1.05}
  \begin{tabular}{@{}llrrrc@{}}
    \toprule
    \textbf{Workload} & \textbf{Suite} & \textbf{MPKI} & \textbf{mem-MPKI} & \textbf{Footprint} & \textbf{Cls} \\
    \midrule
    atax            & PolyBench & 81.4   & 119.6  & 16.0\,MB   & A \\
    mvt               & PolyBench & 56.5   & 83.0   & 16.0\,MB   & A \\
    \midrule
    bicg            & PolyBench & 56.4   & 82.8   & 16.0\,MB   & B \\
    gesummv           & PolyBench & 175.5  & 249.7  & 31.3\,MB   & B \\
    nw                  & Rodinia   & 5.02   & 67.9   & 32.0\,MB   & B \\
    dmr                  & Lonestar  & 2.99   & 18.6   & 24.2\,MB   & B \\
    kmeans              & Rodinia   & 2.09   & 72.5   & 30.3\,MB   & B \\
    sssp           & Lonestar  & 1.58   & 11.3   & 48.0\,MB   & B \\
    mst                & Lonestar  & 0.98   & 6.80   & 73.0\,MB   & B \\
    \midrule
    backprop          & Rodinia   & 0.02   & 0.30   & 9.0\,MB    & --- \\
    bfs                & Rodinia   & 0.03   & 0.20   & 2.4\,MB    & --- \\
    gaussian            & Rodinia   & 0.00   & 0.01   & 0.5\,MB    & --- \\
    hybridsort          & Rodinia   & 0.02   & 0.46   & 12.0\,MB   & --- \\
    lud                 & Rodinia   & 0.03   & 0.58   & 16.0\,MB   & --- \\
    pagerank                  & Pannotia  & 0.06   & 0.40   & 10.9\,MB   & --- \\
    p-bfs                & Parboil   & 0.00   & 0.07   & 1.1\,MB    & --- \\
    mri-gridding         & Parboil   & 0.09   & 6.04   & 226.4\,MB  & --- \\
    sad                       & Parboil   & 0.02   & 1.48   & 8.5\,MB    & --- \\
    sgemm               & Parboil   & 0.00   & 0.07   & 12.0\,MB   & --- \\
    spmv                      & Parboil   & 0.19   & 0.71   & 29.4\,MB   & --- \\
    2mm            & PolyBench & 0.00   & 0.00   & 5.0\,MB    & --- \\
    fdtd2d          & PolyBench & 0.09   & 0.49   & 48.0\,MB   & --- \\
    gramschmidt               & PolyBench & 0.40   & 0.62   & 24.4\,MB   & --- \\
    streamcluster        & Rodinia   & 0.17   & 0.70   & 8.4\,MB    & --- \\
    \bottomrule
  \end{tabular}
\end{table}

\subsection{Evaluated Configurations}

We evaluate six configurations: (i) Baseline: Unmodified LRU replacement, 4\,KB pages, (ii) DEPOT: DEPOT eviction protection enabled with default parameters ($W = 500\text{K}$ cycles, 8192-bit Bloom filter), (iii) 2\,MB baseline: Baseline with 2\,MB huge pages (L2 TLB reconfigured to 128 entries $\times$ 2\,MB, reach = 256\,MB), (iv) 2\,MB+DEPOT: DEPOT with 2\,MB pages, (v) LatPC: State-of-the-art TLB prefetching and compaction~\cite{latpc} enabled, 4\,KB pages, and (vi) LatPC+DEPOT: Both LatPC and DEPOT enabled simultaneously.

\section{Dead-Entry Miss Characterization}
\label{sec:characterization}

\subsection{Dead-Entry Ratio}
\label{sec:char_de_ratio}

\begin{figure}[ht]
  \centering
  \includegraphics[width=\figwidth]{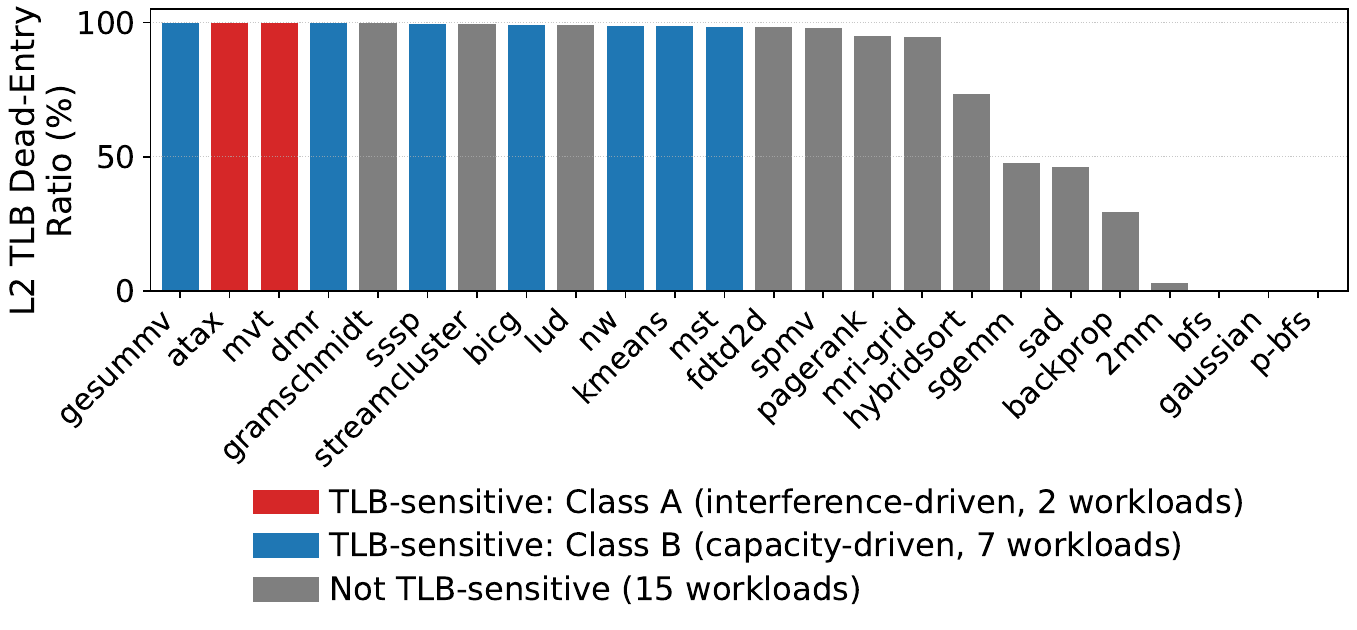}
  \caption{L2 TLB dead-entry ratio for all 24 workloads, sorted descending.}
  \label{fig:de_bars}
\end{figure}

Figure~\ref{fig:de_bars} shows the dead-entry ratio, defined as the fraction of L2 TLB
misses classified as dead-entry re-walks, for all 24 workloads sorted in descending order.
The distribution is striking: dead-entry misses are not a rare edge case but the dominant
miss type across the benchmark suite.
16 of the 24 workloads exhibit dead-entry ratios above 90\%, and all 9 TLB-sensitive workloads exceed 98\% (six of them above 99\%).
This near-unity ratio arises from the interplay between working-set size and access
recurrence.
Once a TLB-sensitive workload reaches steady-state execution, its working set is large
enough to guarantee eviction between successive accesses to the same page, yet small
enough that the same pages are accessed repeatedly within a single kernel invocation.
As a result, virtually every L2 TLB miss encountered in steady state is a re-walk of a
page that the TLB has already translated and discarded.

\noindent Importantly, high DE ratio alone does not imply TLB sensitivity.
Workloads like \texttt{lud} and \texttt{gramschmidt} exceed 99\% DE ratio yet sustain
high throughput because their absolute miss rate is low enough that cumulative stall
time remains negligible, motivating a joint condition of high DE ratio and
high MPKI.

\begin{figure}[t]
  \centering
  \includegraphics[width=\figwidth]{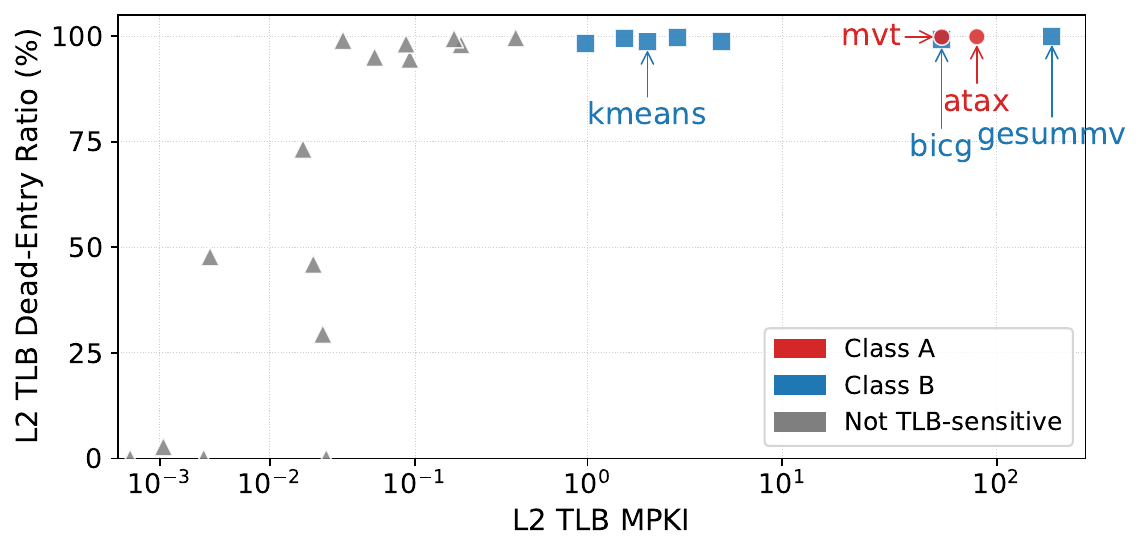}
  \caption{L2 TLB Dead-entry ratio vs.\ L2 TLB MPKI for all 24 workloads.}
  \label{fig:de_mpki}
\end{figure}

Figure~\ref{fig:de_mpki} plots all 24 workloads by DE ratio against L2 TLB MPKI.
The most immediately apparent feature is the vertical gap separating TLB-sensitive
workloads from the rest: 9 sensitive workloads occupy a clearly elevated MPKI
band, while the remaining 15 cluster near the bottom of the MPKI axis regardless
of their DE ratio.
Within the high-MPKI band, all workloads cluster near
100\% dead-entry ratio,
confirming that near-unity DE ratio is a reliable
characteristic of workloads that are both TLB-sensitive and operating in steady state.
In contrast, the lower portion of the plot shows a wide spread of DE ratios among
workloads with low MPKI, ranging from near zero to above 99\%, confirming that DE ratio
alone carries no predictive power for performance sensitivity.

\subsection{Temporal Structure}
\label{sec:char_temporal}

\begin{figure}[ht]
  \centering
  \includegraphics[width=\figwidth]{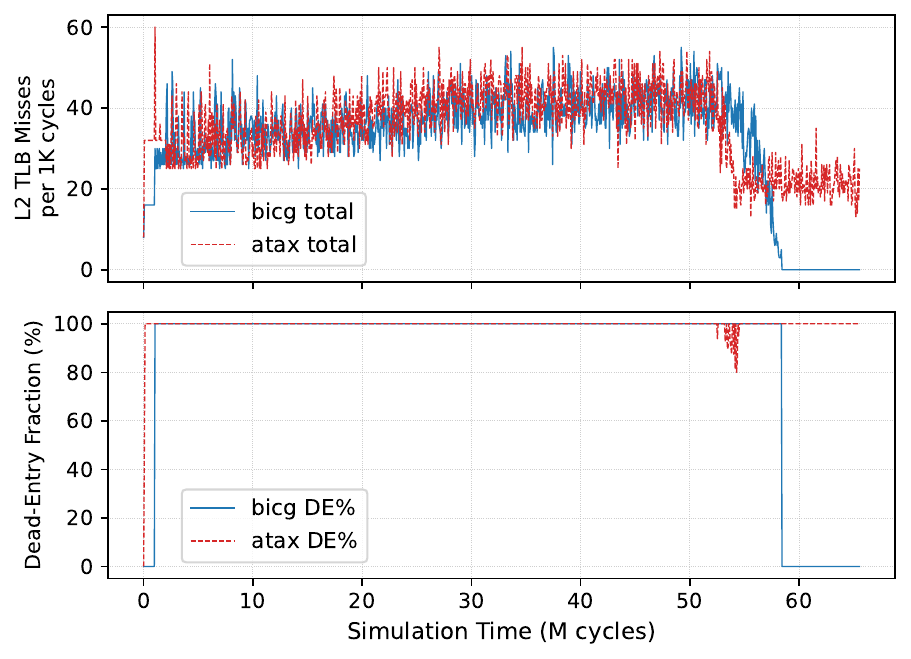}
  \caption{Miss arrival timeseries and dead-entry fraction for \texttt{atax} and \texttt{bicg} in steady state.}
  \label{fig:miss_ts}
\end{figure}

To understand how dead-entry misses evolve over the course of kernel execution,
Figure~\ref{fig:miss_ts} shows the miss arrival timeseries for \texttt{atax} and
\texttt{bicg}, two workloads that share nearly identical DE ratios ($\sim$99.9\%
and $\sim$99.2\%) and comparable MPKI (81 and 56, respectively) yet respond very
differently to eviction-history protection.

\noindent In both workloads, the dead-entry fraction rises sharply from cold-start and stabilizes
near 99\% within the first few thousand kernel cycles.
This rapid convergence shows that the dead-entry phenomenon is a steady-state property
of the workload's memory access pattern rather than a transient warm-up artifact.
After the initial cold-miss period, the TLB is fully populated with the working set
and subsequent misses are almost entirely re-walks of evicted entries.
The miss arrival rate, shown as the total number of misses per time window, is
comparable between the two workloads throughout steady state.
Both issue misses at similar frequencies and both carry the eviction-history signature
on nearly all of them.
The timeseries data alone would lead to the conclusion that the two workloads are
structurally interchangeable, yet their performance responses to the same eviction-history
intervention differ by nearly two orders of magnitude.
Understanding this divergence requires examining beyond miss
rate to how warps share and reuse virtual pages.

\subsection{Access Structure: The Class Discriminator}
\label{sec:char_burstness}

\begin{figure}[ht]
  \centering
  \includegraphics[width=\figwidth]{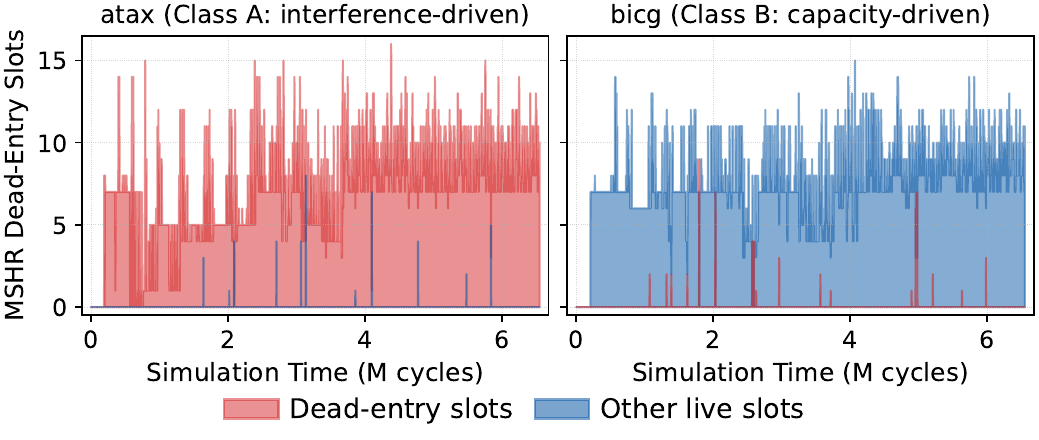}
  \caption{MSHR dead-slot occupancy over time for \texttt{atax} and \texttt{bicg}.}
  \label{fig:burstness}
\end{figure}

The discriminator between \texttt{atax} and \texttt{bicg} is not miss frequency but access structure, specifically whether warps tend to reference the same virtual page or distinct virtual pages at a given moment. Figure~\ref{fig:burstness} shows the MSHR dead-slot occupancy timeseries for both workloads. \texttt{atax} exhibits brief, high-amplitude bursts separated by near-zero occupancy, while \texttt{bicg} shows a flat, sustained occupancy level throughout execution. This difference directly reflects how each workload maps onto the warp execution model.

In \texttt{atax}, each warp computes $y \mathrel{+}= A[i][j] \cdot x[j]$.
All 32 threads within the warp access the same element of the $x[]$ vector, the
same $j$ index, producing a single shared VPN for that warp instruction.
When this VPN is evicted and re-walked, all warps stalled on the same $x[j]$ reference are merged into a single MSHR entry through same-VPN coalescing. A single page walk therefore serves many waiting warps simultaneously, producing the sharp occupancy spikes in Figure~\ref{fig:burstness}. The spike height reflects the number of merged and simultaneously released warps. \texttt{mvt} exhibits the same access pattern for analogous algebraic reasons, making it the second member of this class.

In \texttt{bicg}, each warp computes a column dot product where thread $i$ accesses row $i$ of matrix $A$, generating 32 distinct VPNs per warp instruction with little cross-warp sharing. The MSHR merge depth is therefore approximately one, so each miss requires an independent page walk. Its 2048-by-2048 working set requires approximately 4096 pages, four times the 1024-entry L2 TLB capacity, making evictions fundamentally capacity-driven. Consequently, the MSHR occupancy trace in Figure~\ref{fig:burstness} remains flat and sustained throughout execution because misses form a steady stream of independent capacity evictions rather than burst events. The remaining 6 Class B workloads (\texttt{nw}, \texttt{sssp}, \texttt{kmeans}, \texttt{gesummv}, \texttt{dmr}, and \texttt{mst}) exhibit the same structural property, generating unique VPN streams with working sets that exceed L2 TLB reach.

The two classes can therefore be distinguished by the shape of the MSHR dead-slot
occupancy signal.
Bursty, high-amplitude spikes indicate that warps share virtual pages and that
individual eviction events serve many waiting requestors at once.
Flat, sustained occupancy indicates that each warp accesses distinct pages and that
evictions are driven by working-set capacity that cannot be resolved at the
replacement level.
This temporal signature is observable from existing MSHR occupancy counters without
any offline profiling or application modification, making it a practical runtime
discriminator.

\subsection{Validating Capacity-Driven Misses}
\label{sec:char_2mb}

\begin{figure}[ht]
  \centering
  \includegraphics[width=\figwidth]{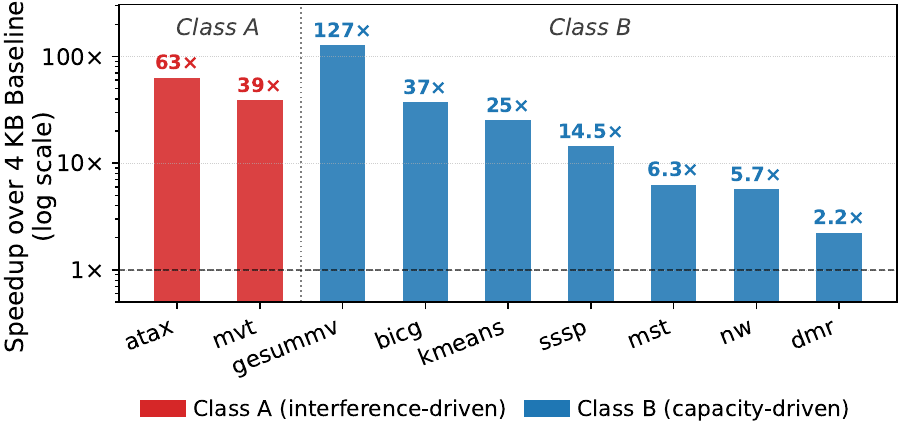}
  \caption{IPC speedup under 2\,MB pages over 4\,KB baseline for all 9 TLB-sensitive workloads.}
  \label{fig:de_2mb}
\end{figure}

The classification of seven workloads as capacity-driven rests on the claim that their
dead-entry misses originate from working-set overflow rather than from interference at
the TLB replacement level.
A direct way to test this claim is to expand TLB reach and observe whether the misses
disappear.
Figure~\ref{fig:de_2mb} shows IPC speedup under 2\,MB huge pages relative to the
4\,KB baseline for all nine TLB-sensitive workloads.
With 2\,MB pages, the L2 TLB reach expands from 4\,MB to 256\,MB, covering a far
larger fraction of each workload's active footprint with the same 1024 entries.

The results validate the capacity-driven diagnosis across all 7 Class~B workloads
through a wide range of speedups that reflect each workload's working-set size relative
to the expanded reach.
\texttt{gesummv} shows the most dramatic response at $128\times$: its 16\,MB working
set fits entirely within the 256\,MB reach, eliminating capacity evictions almost
completely.
\texttt{kmeans} ($25\times$), \texttt{sssp} ($14.5\times$), \texttt{bicg} ($37\times$),
\texttt{nw} ($5.7\times$), and \texttt{mst} ($6.3\times$) each improve by an amount
that tracks how tightly their active footprint is bounded by the reach ceiling.
\texttt{dmr} shows the smallest response at $2.2\times$, reflecting a more irregular
address stream only partially contained within the larger reach.
In every case the improvement confirms that capacity overflow is the root cause.
The two Class~A workloads, \texttt{atax} and \texttt{mvt}, also benefit substantially
from 2\,MB pages, achieving $63\times$ and $39\times$ speedups, respectively.
However, unlike the Class~B workloads, their performance is not fundamentally
constrained by working-set capacity: expanding reach does not eliminate the interference
that causes evictions when many warps compete for the same entries.

\section{DEPOT: Dead-Entry PrOTection}
\label{sec:depot}

\begin{figure}[ht]
  \centering
  \includegraphics[width=\columnwidth]{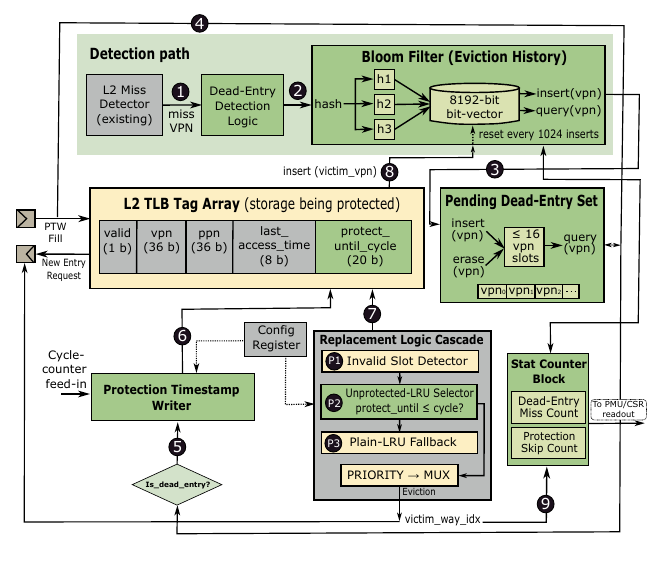}
  \caption{DEPOT mechanism: eviction-history Bloom filter, Pending
  Dead-Entry Set, per-entry protection timer, and protection-aware
  replacement cascade.}
  \label{fig:depot_mech}
\end{figure}

In this section, we propose DEPOT that targets Class~A dead-entry re-walks through a three-phase mechanism
detection, fill, and eviction implemented entirely within the L2 TLB
pipeline. Figure~\ref{fig:depot_mech} shows the hardware. Three additions extend the
existing L2 TLB datapath: an eviction-history Bloom filter, a small Pending
Dead-Entry Set (PDS) that bridges the miss and fill events, and a Protection
Timestamp Writer that drives a new 20-bit per-entry protection-timer field in
the tag array.

\noindent\textbf{Detection (miss path).}
On every L2 TLB miss, the Dead-Entry Detection Logic receives the missed
VPN~\circnum{1} and hashes it through three independent hash functions
($h_1, h_2, h_3$) into an 8192-bit Bloom filter~\cite{bloom_filter} that records recently
evicted VPNs~\circnum{2}.
A filter hit indicates that the missed VPN was recently in the TLB and is
therefore a probable dead-entry re-walk, so the in-flight miss is registered
in the PDS~\circnum{3}-a small ($\le 16$ slot) buffer that holds VPNs
awaiting fill.
Misses whose VPN is absent from the filter (genuine cold or capacity misses
with no eviction history) bypass the PDS and proceed through the standard
page-walk path.

\noindent\textbf{Fill (protection path).}
When the page-table walk completes, the fill triggers a PDS query for the
filling VPN~\circnum{4}.
A PDS hit activates the Protection Timestamp Writer~\circnum{5}, which
reads the cycle counter and sets the newly installed entry's protection
timer to expire at $\textit{now} + W$ cycles~\circnum{6}, where $W$
defaults to 500\,K cycles and is held in a configuration register.
The PDS slot for that VPN is then released.
Fills whose VPN was not previously flagged install with the protection field
untouched and behave identically to baseline.

\noindent\textbf{Eviction (protection-aware replacement).}
On every fill that requires a victim, the Replacement Logic Cascade reads
each entry's protection timer from the tag array~\circnum{7} and selects
through three priority stages:
\emph{\circnum{P1}~Invalid Slot Detector} returns any invalid entry immediately;
\emph{\circnum{P2}~Unprotected-LRU Selector} compares each entry's protection timer
against the current cycle and selects the LRU entry among those whose
protection has expired;
\emph{\circnum{P3}~Plain-LRU Fallback} fires only when every candidate in the set is
still protected, in which case the cascade reverts to standard LRU.
The selected victim's VPN is inserted into the Bloom filter, refreshing the
eviction history~\circnum{8}, and the victim's way index is forwarded to
a small Stat Counter Block~\circnum{9} that records dead-entry-miss and
protection-skip counts.
This graceful degradation is a correctness property: it bounds the worst-case
overhead of DEPOT to zero, since the fallback path is indistinguishable from
baseline LRU. Two operational details keep the mechanism stable across long-running workloads.
First, the filter is reset after every 1024 insertions rather than on a fixed
time interval; this count-based threshold keeps the expected false-positive rate
below 5\% while avoiding the temporal aliasing of a cycle-timer reset that could
misfire across kernel boundaries.
Second, a protection-state flush at each kernel boundary clears all protection
timers, ensuring that a VPN evicted during kernel $K_n$ does not spuriously
protect an entry in kernel $K_{n+1}$, where the same virtual address may map to
a different physical frame after relaunch.

Regarding hardware cost, the Bloom filter occupies 1\,KB of storage.
Each TLB entry requires one additional per-entry protection timer of 20 bits to record
the expiry cycle, yielding $1024 \times 20 = 20\,\text{Kbits} = 2.5\,\text{KB}$ of
per-entry SRAM across the full 1024-entry L2 TLB.
The total overhead is therefore $1\,\text{KB} + 2.5\,\text{KB} = 3.5\,\text{KB}$,
amortized across the existing TLB array. DEPOT's eviction-history Bloom filter is structurally similar to the evicted-address
filter introduced by Seshadri et al.~\cite{evicted_addr_filter} for cache pollution and
thrashing mitigation; we apply the same primitive at the TLB layer, where eviction-history
information drives protection rather than insertion policy.

\section{Performance Evaluation and Analysis}
\label{sec:evaluation}

\subsection{DEPOT Performance}
\label{sec:o3_real}

\begin{figure}[t]
  \centering
  \includegraphics[width=\figwidth]{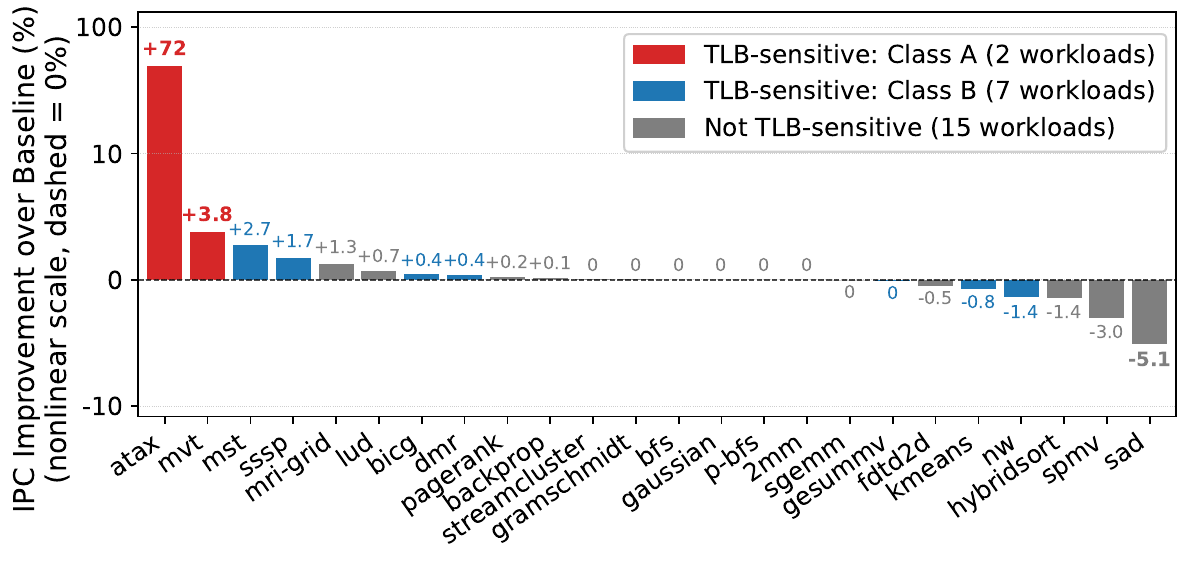}
  \caption{IPC improvement of DEPOT over 4\,KB baseline for all 24 workloads, sorted descending.}
  \label{fig:o3_gains_bar}
\end{figure}

Figure~\ref{fig:o3_gains_bar} shows IPC improvement of DEPOT for all 24 workloads.
The results divide cleanly along the class boundaries established in
Section~\ref{sec:characterization}.
Among the 15 not-TLB-sensitive workloads, IPC improvement is near zero throughout, with
occasional small negative values including \texttt{sad} at $-5.1\%$ and \texttt{spmv}
at $-2.9\%$.
These minor regressions arise because DEPOT occasionally fires on workloads with
moderate DE ratios, marking re-installed entries as protected even though TLB misses
are not on their critical path; with MPKI of 0.02 and 0.19 respectively,
\texttt{sad} and \texttt{spmv} incur negligible TLB stall time at baseline,
so the observed losses reflect second-order LRU displacement rather than
translation overhead.
The LRU fallback bounds worst-case degradation but activates only when all entries
in a set are simultaneously protected, so partial-protection sets can still make
suboptimal eviction choices, which accounts for the small residual regression.
Among the two Class~A workloads, \texttt{atax} improves from 0.354 to 0.610 IPC for
a gain of $+72\%$, reflecting near-complete elimination of dead-entry re-walk stalls
that previously dominated its execution time.
\texttt{mvt} improves from 0.574 to 0.596 IPC for a gain of $+3.8\%$.
The smaller gain on \texttt{mvt} reflects its peak MSHR merge depth of 15 versus
\texttt{atax}'s 17 and its higher baseline IPC, indicating that TLB stalls already
represent a smaller fraction of its execution time before DEPOT is applied.
Both workloads converge to similar post-DEPOT IPC values near 0.60, consistent with
reaching a shared performance ceiling determined by compute throughput and memory
bandwidth rather than translation overhead.
All seven Class~B workloads show near-zero DEPOT gain, ranging from $-1.4\%$ to
$+2.7\%$, within simulation noise. This result holds across PolyBench, Lonestar, and Rodinia workloads: when each warp generates distinct VPNs, protecting one re-installed entry is immediately offset by another capacity eviction, while the LRU fallback ensures zero overhead.

\begin{figure}[ht]
  \centering
  \includegraphics[width=\figwidth]{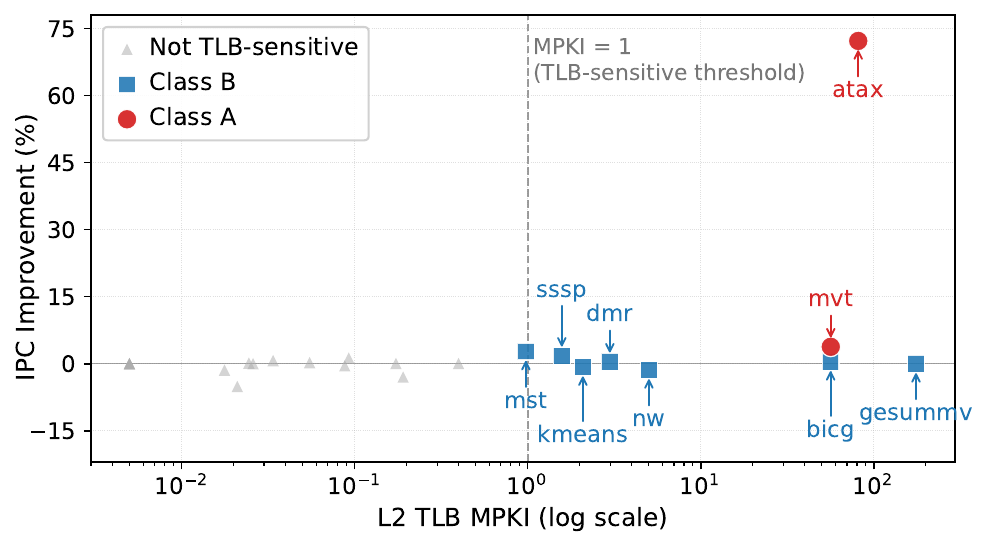}
  \caption{IPC improvement of DEPOT vs.\ L2 TLB MPKI for all 24 workloads.}
  \label{fig:o3_scatter}
\end{figure}

Figure~\ref{fig:o3_scatter} plots IPC improvement of DEPOT against MPKI for all 24 workloads,
revealing a clean two-regime structure.
Workloads with MPKI below 1 cluster near zero gain regardless of their dead-entry ratio,
confirming that TLB miss rate is a necessary condition for sensitivity.
Within the high-MPKI group, workloads split by access structure: those with shared VPN
access show positive gains, while those with unique VPN access per warp show near-zero
gains.
The scatter plot makes visible what the aggregate bar chart in
Figure~\ref{fig:o3_gains_bar} obscures: MPKI identifies workloads for which TLB
behavior matters at all, while access structure determines whether eviction-history
protection can improve it.
To bound the overhead from false positives, we also evaluate a worst-case scenario with
a fully saturated Bloom filter, where every miss query returns a positive response
regardless of actual eviction history.
Under saturation, \texttt{atax} IPC remains at 0.610, identical to the normal DEPOT
result.
False positives cause DEPOT to protect entries that did not originate from dead-entry
re-walks, but the LRU fallback absorbs the resulting mismatch without measurable
performance impact, showing that the mechanism is tolerant of false positive pressure.

\begin{figure}[ht]
  \centering
  \includegraphics[width=\figwidth]{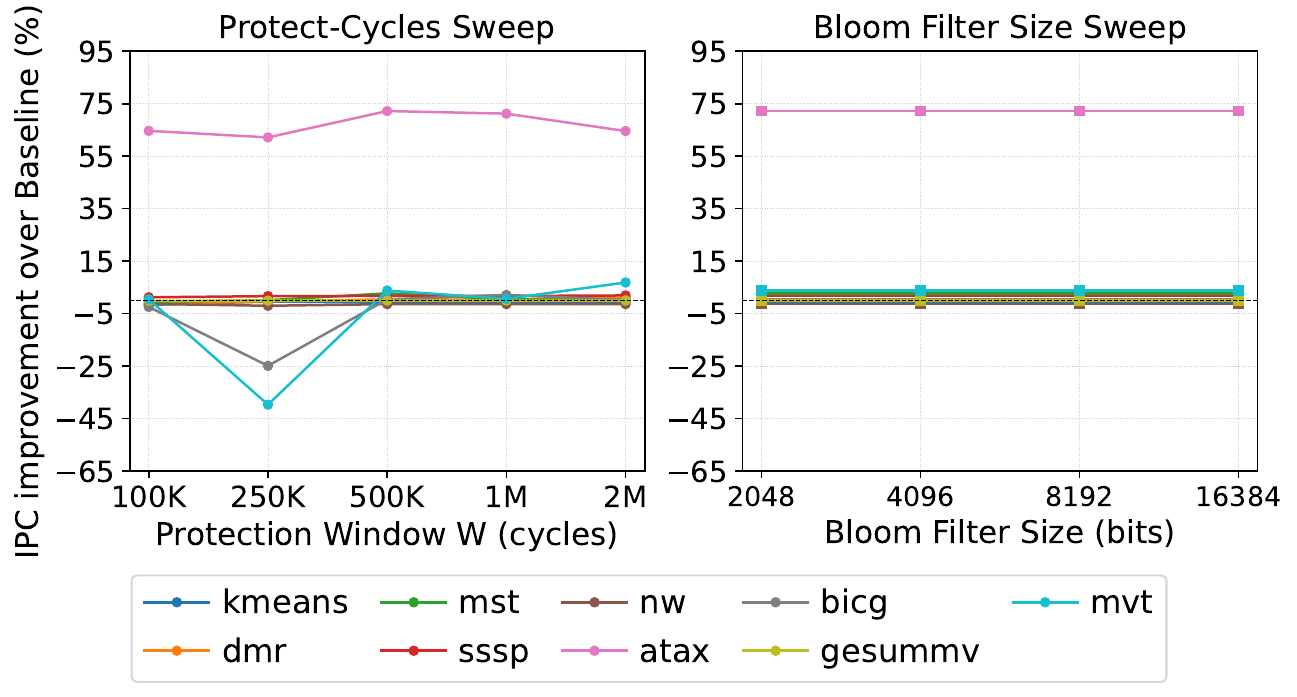}
  \caption{DEPOT parameter sensitivity for 9 TLB-sensitive workloads: protection window $W$ (left) and Bloom filter size (right).}
  \label{fig:params}
\end{figure}

Figure~\ref{fig:params} shows DEPOT sensitivity to its two configurable parameters for 9 TLB-sensitive workloads. Both the protection-window sweep (100\,K to 2\,M) and Bloom-filter sweep (2048 to 16384 bits) remain nearly flat across the tested range. This indicates that TLB-sensitive workloads experience re-walk bursts within timescales covered by any reasonable protection window, while the eviction stream remains well below the saturation threshold of even the smallest Bloom filter. DEPOT is therefore robust to parameter choice, and the default configuration of a 500\,K-cycle protection window and an 8192-bit Bloom filter is effective across all evaluated workloads without tuning.

\subsection{Class~A/B Asymmetry}
\label{sec:disc_classes}
The performance asymmetry between Class A and Class B follows directly from the access-structure analysis in Section~\ref{sec:char_burstness} and explains both the large gains on \texttt{atax} and the near-zero results on the Class B workloads. In Class A workloads, a single dead-entry eviction is amplified through MSHR merging because many warps simultaneously reference the same virtual page. When DEPOT protects the re-installed entry, it eliminates both the initial re-walk and the merged re-walks that would otherwise stall many warps simultaneously. On SM86, page-walk latency is approximately 300–500 cycles, and releasing dozens of stalled warps with a single protection event can recover thousands of cycles, consistent with the observed +72\% IPC improvement.

In Class B workloads, by contrast, the TLB is continuously occupied by a working set that exceeds TLB reach, and each warp accesses distinct pages with little cross-warp sharing. Protecting one re-installed entry therefore only shifts the next eviction to another entry without reducing the overall eviction rate. Because MSHR merging is minimal, each page walk serves only one warp, so the benefit of protection is limited. When all entries are under protection pressure, DEPOT falls back to standard LRU replacement, ensuring negligible overhead in this regime.

\subsection{Composition with LatPC}
\label{sec:o3_latpc}

LatPC combines stride-based prefetching with entry compaction and represents the
state-of-the-art in GPU TLB optimization.
It trains a stride predictor for each TLB entry and issues prefetch requests before
demand misses occur, targeting cold and capacity-driven misses through access-pattern
prediction.
DEPOT and LatPC address fundamentally different sources of TLB misses: LatPC reduces
misses by predicting future accesses and filling the TLB ahead of demand, while DEPOT
reduces misses by preventing recently installed entries from being prematurely evicted
at the replacement boundary.
Because their mechanisms operate on different points in the TLB pipeline, one does not
subsume the other: on interference-driven workloads the two compose additively, while
on capacity-saturated workloads they can contend, as the results below show.

\begin{figure}[ht]
  \centering
  \includegraphics[width=\figwidth]{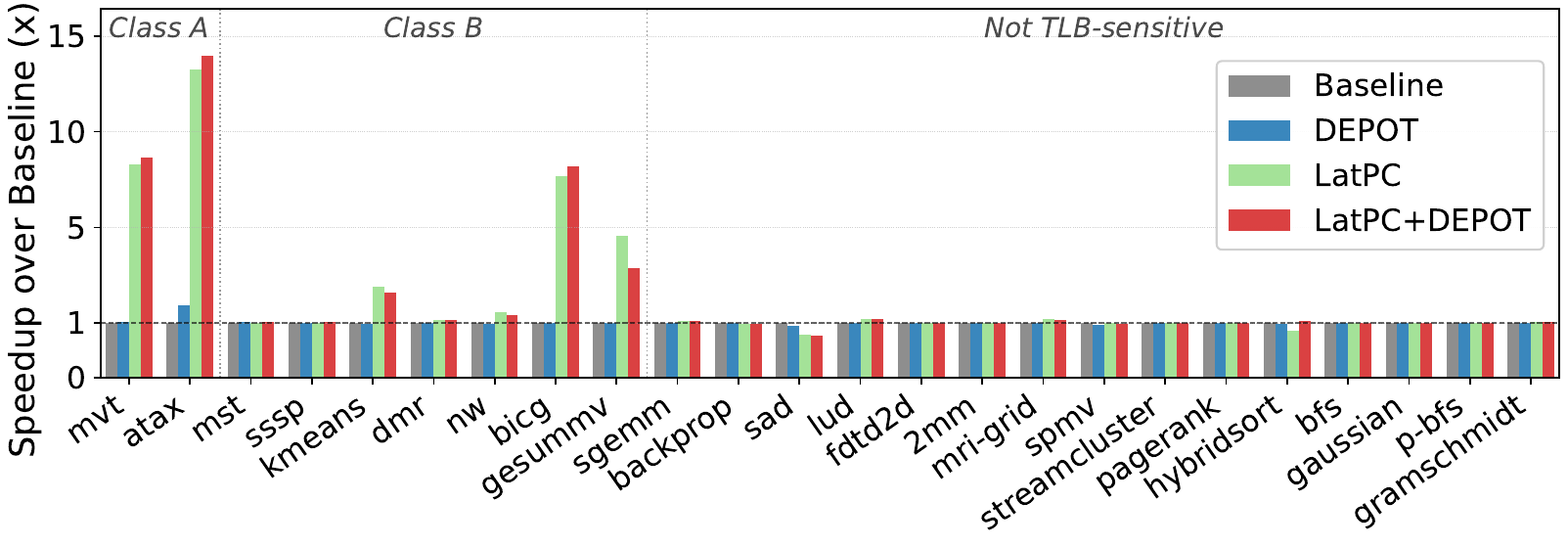}
  \caption{IPC comparison across four configurations for all 24 workloads.}
  \label{fig:ipc}
\end{figure}

Figure~\ref{fig:ipc} shows the composition result across all configurations.
LatPC alone achieves large gains on TLB-sensitive workloads, with \texttt{atax}
improving by $+1227\%$ from 0.354 to 4.70\,IPC, \texttt{bicg} by $+667\%$ from 0.601
to 4.61\,IPC, and \texttt{mvt} by $+729\%$ from 0.574 to 4.76\,IPC.
These gains reflect LatPC's elimination of the dominant cold-miss bottleneck, which
accounts for the large majority of TLB stalls in these workloads.
LatPC+DEPOT outperforms LatPC alone on the Class~A workloads, with \texttt{atax}
improving further from 4.70 to 4.95\,IPC ($+5.4\%$) and \texttt{mvt} from 4.76 to
4.95\,IPC ($+4.0\%$).
Across the Class~B workloads, LatPC+DEPOT shows mixed results: \texttt{bicg}
($+6.9\%$), \texttt{mst} ($+2.5\%$), and \texttt{sssp} ($+2.1\%$) gain modestly,
\texttt{dmr} is negligible ($-0.9\%$), while \texttt{kmeans} ($-8.3\%$),
\texttt{nw} ($-10.0\%$), and \texttt{gesummv} ($-28.5\%$) regress.
These regressions arise because DEPOT's eviction-protection policy conflicts with
LatPC's prefetch-driven replacement in capacity-saturated workloads: LatPC must
freely evict entries to install ahead-of-demand translations, and DEPOT's protected
entries block those slots, delaying prefetch fills and causing net IPC loss.

The additive improvement on Class~A workloads arises because LatPC does not eliminate all
dead-entry re-walks even when its prefetcher is operating correctly.
The key reason is the asymmetry between what the prefetcher targets and what eviction
produces.
LatPC observes that a warp accesses VPN $A$, then predicts and prefetches the next
VPN in the stride sequence.
While the prefetch is in flight, VPN $A$ itself may be displaced from the TLB by LRU
replacement, because the prefetch installs a new entry and the LRU policy does not
know that $A$ will be accessed again at the next stride period.
When that next access to $A$ arrives it finds the entry gone, a dead-entry miss that
LatPC's forward-looking predictor cannot catch; DEPOT observes the eviction of $A$
at the replacement boundary and protects the re-installed entry, covering exactly
this residual class.
Where the combination is beneficial---the Class~A workloads and \texttt{bicg}---the
2 to 7\% additional IPC improvement reflects how frequently this interaction between
prefetch-driven installation and LRU displacement produces residual dead-entry events.
The marginal hardware cost of adding DEPOT to a LatPC-equipped TLB is a 1\,KB Bloom
filter, negligible relative to LatPC's per-entry stride prediction tables, making the
combination an efficient pairing for workloads that benefit from both.

\section{Discussion and Limitations}
\label{sec:discussion}

The 2\,MB page results in Section~\ref{sec:char_2mb} raise a natural question about whether
DEPOT is necessary if larger pages already resolve TLB pressure.
The two mechanisms address different root causes and are not interchangeable.
Larger pages expand TLB reach and directly relieve capacity eviction in Class~B workloads,
but they do not eliminate interference-driven eviction in Class~A workloads, nor are they
always available: GPU workloads frequently use custom memory pools or third-party libraries
that do not guarantee 2\,MB alignment, and virtualized environments may not expose huge page
support to guest workloads at all.
DEPOT operates entirely within the TLB microarchitecture and is transparent to the OS,
driver, and application, making it applicable in settings where large pages are unavailable
or impractical.

DEPOT targets interference-driven dead-entry misses and provides no benefit for workloads
whose TLB pressure arises from working-set overflow beyond the TLB reach.
The Class~B workloads in this study fall squarely into this regime, and the near-zero DEPOT
gains on those workloads confirm that eviction-history protection does not substitute for
additional TLB capacity or larger page mappings.
Practitioners diagnosing TLB performance issues should therefore use the two-class
characterization framework to identify the dominant root cause before selecting an
intervention, as applying DEPOT to a capacity-driven workload not only fails to improve performance
but can actively regress it when a prefetcher such as LatPC is also present,
by blocking the replacement slots that prefetch fills require.

Composition with an aggressive prefetcher is a further limitation.
DEPOT applied alone never regresses, but composing it with LatPC is not universally
additive: on three capacity-driven Class~B workloads---\texttt{kmeans}, \texttt{nw},
and \texttt{gesummv}---LatPC+DEPOT regresses relative to LatPC alone (by $8.3\%$,
$10.0\%$, and $28.5\%$ respectively), as DEPOT's protection window and LatPC's
prefetch installs contend for L2 TLB capacity.
This is consistent with the taxonomy rather than a counterexample to it: DEPOT is an
interference-driven (Class~A) optimization, and these regressions occur only when it
is misapplied to capacity-driven workloads atop a prefetcher.
A capacity-aware coupling of the two mechanisms---for instance, suspending protection
while the prefetcher is actively installing---is a worthwhile direction for future work.

The evaluation in this paper is conducted on the SM86 Ampere microarchitecture using a
validated simulation model.
The qualitative behavior of dead-entry misses, including MSHR-merge amplification and the
distinction between interference-driven and capacity-driven eviction, is expected to
generalize across GPU microarchitectures that share the same L1 and L2 TLB organization.
However, the specific quantitative thresholds used to classify workloads and the absolute
IPC improvement reported here are calibrated to SM86 parameters such as the 1024-entry L2 TLB
capacity and 16 hardware page-table walkers, and may differ on architectures with
substantially different TLB configurations.

\section{Conclusion}
\label{sec:conclusion}

Dead-entry TLB misses are a structurally distinct, avoidable miss class that accounts
for over 99\% of L2 TLB misses in the most TLB-sensitive GPU workloads.
We characterize dead-entry misses across 24 GPU workloads, using MPKI $\geq 1$ as a
TLB-sensitivity threshold to identify 9 TLB-sensitive workloads and sub-classify them
by access structure into Class~A and Class~B,
with Class~B membership validated by 2\,MB IPC experiments confirming reach-limited
root causes.
DEPOT, a 1\,KB Bloom-filter eviction protection mechanism, eliminates Class~A
dead-entry re-walks with up to 72\% IPC improvement, degrades gracefully to zero
overhead on Class~B, and composes additively with LatPC~\cite{latpc} on
interference-driven workloads to deliver 2 to 7\% further improvement at negligible
marginal hardware cost, pushing \texttt{atax} from 4.70 to 4.95\,IPC for a combined
$14\times$ improvement over the 4\,KB baseline.
Several directions remain for future work.
The temporal MSHR occupancy signatures identified in this paper suggest the potential
for runtime-adaptive TLB policies that dynamically distinguish interference-driven from
capacity-driven miss behavior.
In addition, integrating eviction-history signals with TLB-aware warp scheduling or
page-placement mechanisms may further reduce destructive translation interference
before evictions occur.

\bibliographystyle{IEEEtranS}
\bibliography{reference}

\end{document}